\documentclass[prl,floatfix,preprint,superscriptaddress,showpacs]{revtex4}
\usepackage{graphicx}% Include figure files
\usepackage[english]{babel}
\usepackage[usenames]{color}
\usepackage{amsmath,amsfonts,amssymb,float}
\usepackage{dcolumn}% Align table columns on decimal point
\usepackage{bm}% bold math
\def\clf{Central Laser Facility, STFC Rutherford Appleton Laboratory, Didcot, OX11 0QX, United Kingdom}

\def\strathclyde{SUPA, Department of Physics, University of Strathclyde, Glasgow, G4 0NG, United Kingdom}
\def\ist{GoLP/Instituto de Plasmas e Fus\~ao Nuclear, Instituto Superior T\'ecnico, Universidade de Lisboa, 1049-001 Lisbon, Portugal}
\def\llnl{Lawrence Livermore National Laboratory, Livermore, California, USA}
\def\standrews{University of St Andrews, St Andrews, Fife KY16 9AJ, United Kingdom}
\def\oxford{Department of Physics, University of Oxford, Oxford OX1 3PU, UK}
\def\dcti{DCTI/ISCTE Lisbon University Institute, 1649-026 Lisbon, Portugal}
\def\voetnoot{Authors E. Alves and R. Trines contributed equally to this work.}

\begin{document}
\title{A robust plasma-based laser amplifier via stimulated Brillouin
scattering}
\author{E.P. Alves}
\altaffiliation{\voetnoot}
\affiliation{\ist}
\author{R.M.G.M. Trines}
\altaffiliation{\voetnoot}
\affiliation{\clf}
\author{K.A. Humphrey}
\affiliation{\strathclyde}
\author{R. Bingham}
\affiliation{\clf}
\affiliation{\strathclyde}
\author{R.A. Cairns}
\affiliation{\standrews}
\author{F. Fi\'uza}
\affiliation{\llnl}
\author{R.A. Fonseca}
\affiliation{\ist}
\affiliation{\dcti}
\author{L.O. Silva}
\affiliation{\ist}
\author{P.A. Norreys}
\affiliation{\oxford}
\affiliation{\clf}
\date\today

\begin{abstract}
It is shown here that Brillouin amplification can be used to produce
picosecond pulses of petawatt power. Brillouin amplification is far
more resilient to fluctuations in the laser and plasma parameters than
Raman amplification, making it an attractive alternative to Raman
amplification. Through analytic theory and multi-dimensional computer
simulations, a novel, well-defined parameter regime has been found,
distinct from that of Raman amplification, where pump-to-probe
compression ratios of up to 100 and peak laser fluences over 1
kJ/cm$^2$ with 30\% efficiency have been achieved. High pulse quality
has been maintained through control of parasitic instabilities.
\end{abstract}
\pacs{52.38.-r, 42.65.Re, 52.38.Bv, 52.38.Hb}
\maketitle

Amplification of laser beams via parametric instabilities in plasma
(stimulated Raman and Brillouin scattering) has been proposed a number
of times \cite{maier66,milroy77,milroy79, capjack82,andreev89}, but
came into its own only relatively recently
\cite{shvets99,kirkwood99,ping04,weber1,ren07,ping09,lancia,trines10,
  kirkwood11,trines11,toroker12}. Brillouin scattering has also been
used to transfer energy via the Cross-Beam Energy Transfer scheme at
the National Ignition Facility
\cite{kruer96,williams04,michel09,glenzer10,michel10,hinkel11,moody12}.
Both Raman and Brillouin scattering have been studied extensively in
the context of Inertial Confinement Fusion \cite{tanaka82,walsh84,
  vill87,mori94,langdon02,lindl04,hinkel05,froula07,michel11,glenzer11};
Raman scattering also in the context of wakefield acceleration
\cite{forslund85,decker94,rousseaux95,karl95,tzeng96,decker96,moore97,
  tzeng98,gordon98,matsuoka10}. Raman and Brillouin scattering are
processes where two electromagnetic waves at slightly different
frequencies propagating in plasma exchange energy via a plasma
wave. For Raman scattering, this is a fast electron plasma wave, while
for Brillouin scattering it is a slower ion-acoustic wave
\cite{forslund}. When it comes to laser beam amplification, Raman and
Brillouin scattering have different properties and serve different
purposes. Raman amplification yields the shortest output pulses and
the highest amplification ratios, but it is sensitive to fluctuations
in the experimental parameters and requires high accuracy in the
matching of laser and plasma frequencies. Brillouin amplification
yields lower peak intensities or amplification ratios, but is far more
robust to parameter fluctuations or frequency mismatch, more efficient
(as less laser energy stays behind in the plasma wave) and more
suitable for the production of pulses with a high total power or
energy.

For both Raman and Brillouin amplification, two main goals can be
identified: first, maximising the final power and energy content of
the pumped pulse, and second, ensuring that the pumped pulse has the
best possible quality, i.e. a smooth envelope and a high contrast
(low-intensity pre-pulse). Production of kilojoule, picosecond laser
pulses of good quality using Raman amplification has been explored by
Trines \emph{et al.}  \cite{trines10,trines11}. Here it will be shown
that a similar approach also works for Brillouin amplification in the
so-called ``strong coupling'' regime. The lower compression ratios
obtained for Brillouin (as compared to Raman) amplification work in
favour of this scheme for the production of high-energy picosecond
pulses: higher pump intensities can be used to obtain a given probe
duration, allowing the use of smaller diameters of the pulses and the
plasma column.

To explore how the final duration of a Brillouin-amplified probe pulse
can be controlled, we use the self-similar model of Andreev \emph{et
  al.}  \cite{weber1}. We start from a homogeneous plasma with
electron number density $n_0$, plasma frequency $\omega_p^2 = e^2
n_0/(\varepsilon_0 m_e)$, ion plasma frequency $\omega_{pi} =
\omega_p\sqrt{Z^2 m_e/m_i}$, electron/ion temperatures $T_e$ and
$T_i$, electron thermal speed $v_T^2 = k_B T_e/m_e$, Debye length
$\lambda_D = v_T/\omega_p$, and a pump laser pulse with wave length
$\lambda$, intensity $I$, frequency $\omega_0 = 2\pi c/\lambda$,
dimensionless amplitude $a_0 \equiv 8.55\times 10^{-10} \sqrt{g}
\sqrt{I \lambda^2 [\mathrm{Wcm}^{-2}\mu \mathrm{m}^2]}$, where $g=1$
($g=1/2$) denotes linear (circular) polarisation, and wave group speed
$v_g/c = \sqrt{1-\omega_p^2/\omega_0^2} = \sqrt{1-n_0/n_{cr}}$. Let
the durations of pump and probe pulse be given by $\tau_{pu}$ and
$\tau_{pr}$, and define $\gamma_B = (\sqrt{3}/2) [a_0 (v_g/c)
  \omega_{pi} \sqrt\omega_0]^{2/3}$, the Brillouin scattering growth
rate in the strong-coupling regime \cite{forslund}. Then a full
expansion of the self-similar coordinate $\xi$ of Ref. \cite{weber1}
yields:
\begin{equation}
\label{eq:ssbril}
a_0(v_g/c) \omega_{pi} \tau_{pr} \sqrt{\omega_0 \tau_{pu}} = \sqrt{2g/\eta} \xi_B,
%\gamma_B\tau_{pr} \sqrt{\gamma_B\tau_{pu}} = \xi_B \sqrt{g/\eta},
\end{equation}
where $\xi_B \approx 3.5$ is a numerical constant and $\eta$ denotes
the pump depletion efficiency. The physical interpretation of this
expression is that the duration of the probe pulse is similar to the
time it takes the probe to deplete the counterpropagating pump: for
increasing probe amplification (i.e. longer $\tau_{pu}$) or pump
intensity, pump depletion is more rapid and $\tau_{pr}$ decreases.
This allows one to tune the final probe duration via the properties of
the pump beam, similar to Raman amplification \cite{trines11}.

Using the energy balance $a^2_{pr} \tau_{pr} = \eta a_0^2 \tau_{pu}$,
we also find a relation between amplitude and duration of the growing
probe pulse:
\begin{equation}
\label{eq:topt}
a_{pr}^2 \tau_{pr}^3 = 2g\xi_B^2 [\omega_{pi}^2 \omega_0
  (1-\omega_{pe}^2/\omega_0^2)]^{-1}.
\end{equation}
We repeat this process for Raman amplification to obtain a similar
relation: applying the same energy balance to the Raman self-similar
equation $a_0^2 \omega_0 \omega_p \tau_{pu}\tau_{pr} = (2g/\eta)
\xi_M^2$, we find $a_{pr} \tau_{pr} = \sqrt{2g} \xi_M/\sqrt{\omega_0
  \omega_{pe}}$ with $\xi_M \sim 5$ for a Raman-amplified pulse. This
means that the initial probe pulse duration is not a free parameter:
Eq. (\ref{eq:topt}) dictates the optimal initial probe pulse duration
$\tau_{opt}$ for a given initial probe pulse amplitude $a_1$.  From
previous numerical work on Raman \cite{kim03,trines11} and Brillouin
amplification \cite{lehmann13a,lehmann13b}, it follows that if the
probe pulse is too short for its amplitude initially, it will first
generate a much longer secondary probe pulse behind the original probe
[which does fulfill Eq. (\ref{eq:topt})] and this secondary probe will
then amplify while the original short probe will hardly gain in
intensity. Thus, trying to produce ultra-short laser pulses via
Brillouin amplification by reducing the initial pulse duration simply
does not work. Earlier attempts in this direction
\cite{weber13,riconda13} showed no increase in total pulse power (as
opposed to pulse peak intensity), confirming the results of
Ref. \cite{lehmann13a,lehmann13b}.

\begin{figure*}[ht]
\includegraphics[width=0.8\textwidth]{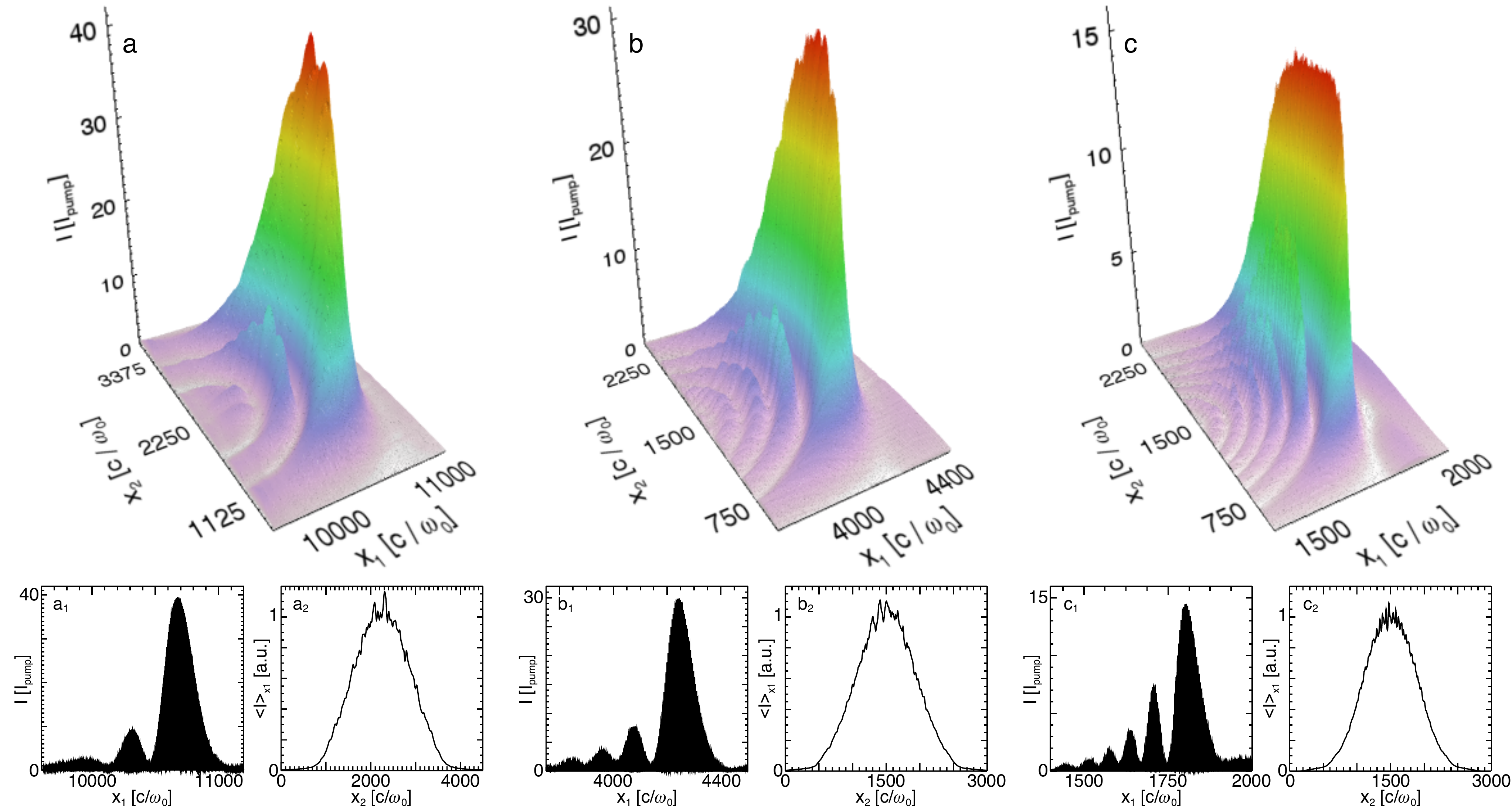}
\caption{Brillouin-amplified probe pulses for pump/probe intensities
  of a) $10^{14}$, b) $10^{15}$ and c) $10^{16}$ W cm$^{-2}$ for
  $n_0/n_{cr} = 0.3$. Pump pulse durations are 11.4 ps, 3.8 ps and 1.1
  ps respectively. The 3D visualizations illustrate the amplified
  probe pulses at 10\% filamentation level. Frames a$_1$--c$_1$ show the
  longitudinal intensity profile taken at the center of the probe, and
  frames a$_2$--c$_2$ show the average transverse intensity profile along
  the longitudinal direction normalized to the average peak
  intensity.}
\label{fig:1}
\end{figure*}

To further investigate Brillouin amplification, in particular limiting
factors such as filamentation and wave breaking of the ion wave, we
have carried out a sequence of particle-in-cell (PIC) simulations
using OSIRIS \cite{osiris1,osiris2,osiris3}. Parameters varied in
these simulations are the pump intensity ($I_0 = 10^{14}$, $10^{15}$
or $10^{16}$ W cm$^{-2}$) and the interaction length. The laser wave
length was $\lambda = 1\ \mu$m and the plasma density was set at
$n_0/n_{cr} = 0.3$, to eliminate parasitic Raman scattering. Such
scattering can do great damage to the envelope of the amplified pulse,
as discussed below. The ion-electron mass ratio was $m_p/m_e = 1836$
and $T_e=T_i = 500$ eV. The initial probe pulse intensity was chosen
to be the same as the pump intensity, and the initial probe duration
was half the value predicted by (\ref{eq:topt}), because this yielded
a somewhat better performance.  The plasma column was given a constant
density, while the plasma length was determined dynamically as these
simulations were conducted in a moving window with the pump pulse
implemented as a boundary condition on the leading edge
\cite{mardahl_mw}.

We have performed two-dimensional moving window simulations, using a
spatial resolution of $dx = \lambda_D/2$ and $dy=0.5c/\omega_0$, with
25 particles per cell per species and quadratic interpolation for the
current deposition. Collisions were not included in the simulations:
while collisions do induce an intensity threshold on both Brillouin
and Raman scattering, the intensities we use are too far above that
for collisions to make much of a difference. Both pump and probe pulses
have identical transverse Gaussian envelopes, with
waist sizes ($W_0$) of $W_0 = 1000c/\omega_0 = 160 \mu$m for
the $10^{15}$ and $10^{16}$ W cm$^{-2}$ scenarios, and $W_0 =
1500c/\omega_0 = 240 \mu$m for the $10^{14}$ W cm$^{-2}$ scenario; these focal
spots are chosen to be wide enough to contain $> 6$ filamentation
wavelengths at their respective initial intensity. The probe pulses
have $\sin^2$ temporal profiles, with durations corresponding to
$\tau_1 = \tau_\mathrm{opt}/2$ determined from (\ref{eq:topt}). The
pump pulses have a flat temporal profile with a short rise time of
$500\ \omega_0^{-1} \simeq 260$ fs.

\begin{figure*}[ht]
\includegraphics[width=0.6\textwidth]{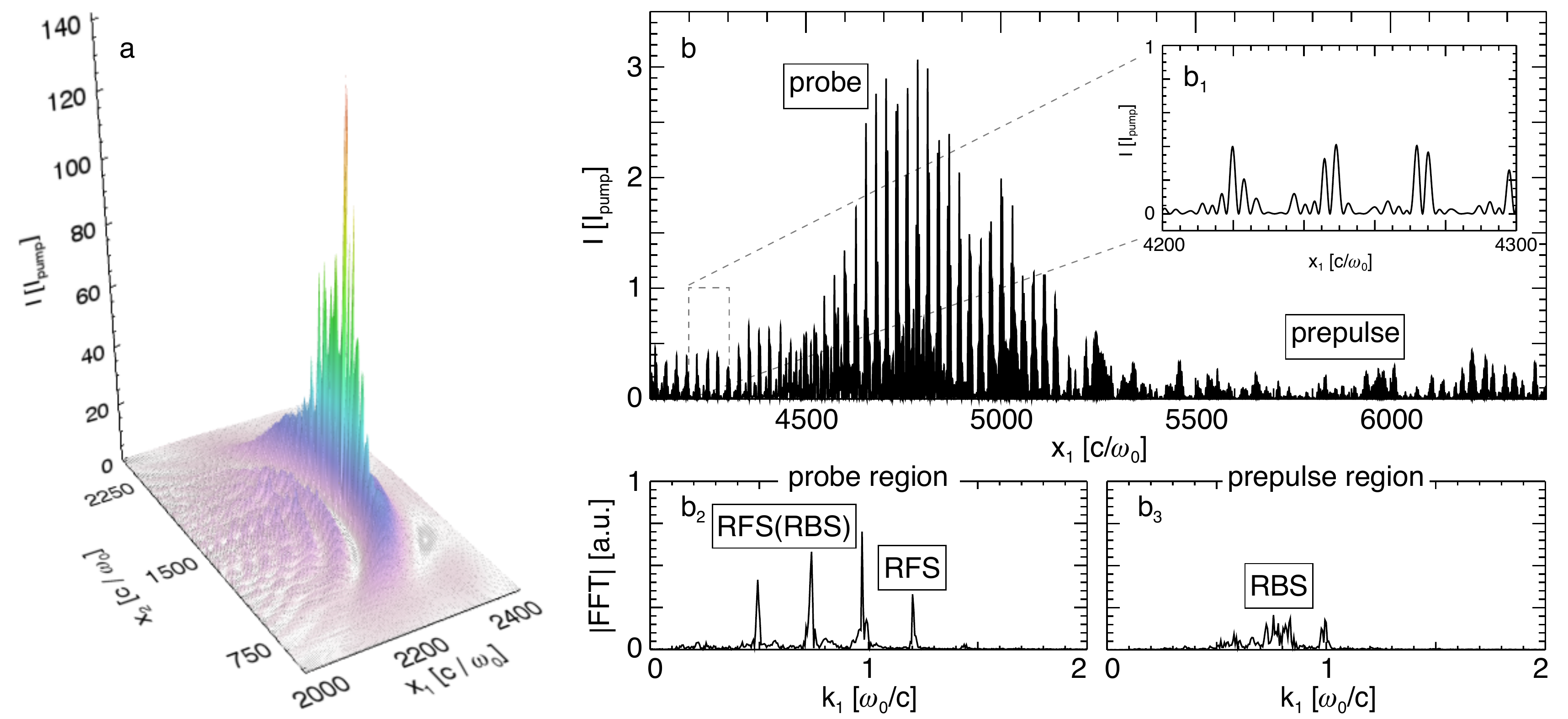}
\ \includegraphics[width=0.3\textwidth]{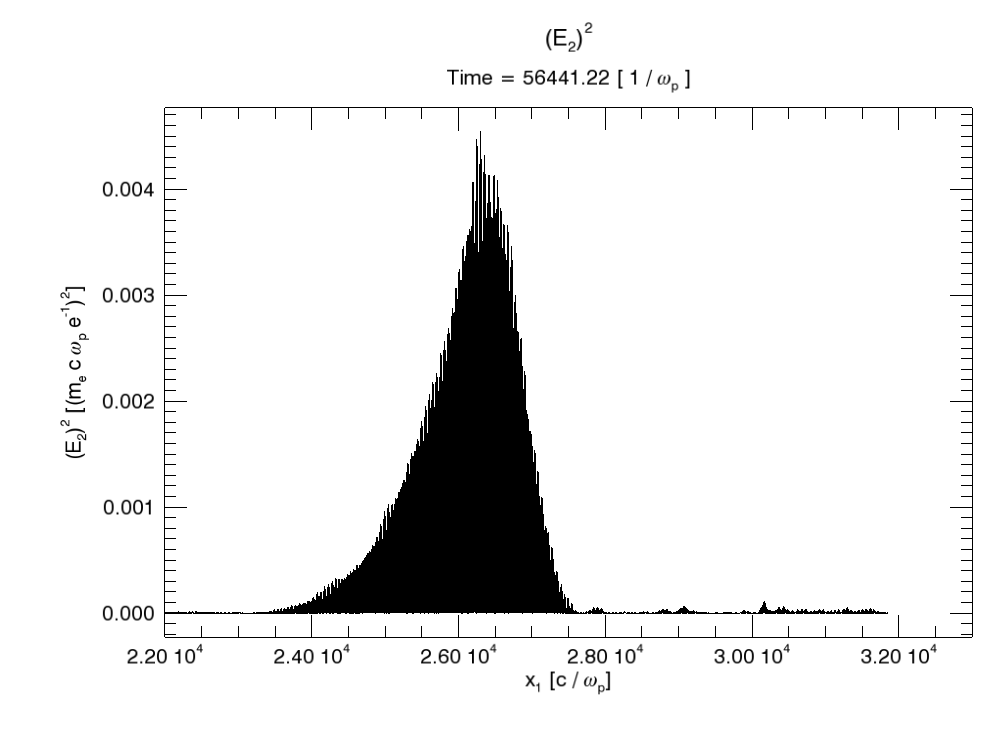}
\caption{Main parasitic instabilities associated with Brillouin
  amplification in a) over-quarter-critical ($n_0/n_{cr}=0.3$) and b)
  sub-quarter-critical ($n_0/n_{cr}=0.05$) density regimes. Examples
  are shown for pump/probe intensities of $10^{16}$ W
  cm$^{-2}$. Distortion of the probe's transverse intensity profile
  due to filamentation is shown in a). Pump-induced RBS/RFS and
  probe-induced RFS are shown in b); inset b$_1$ reveals the
  development of incoherence at the probe tail, and insets b$_2$ and
  b$_3$ show the spectral signatures of the probe and prepulse
  regions, respectively. Frame c) shows the amplified probe for
  $n_0/n_{cr}=0.01$ and pulse intensities of $10^{15}$ W cm$^{-2}$. For
  this case, the pulse envelope is significantly smoother and the
  prepulse intensity much lower, even though the interaction time is
  five times longer than in frame b).}
\label{fig:2}
\end{figure*}

For $n_0/n_{cr} = 0.3$ there will be no Raman backscattering from
noise by the pump, i.e. no significant prepulse, and no modulation of
the probe pulse envelope by Raman forward scattering. Thus, transverse
filamentation of the probe pulse becomes the limiting factor for
amplification, while self-focusing and wave breaking are found to be
insignificant. The interaction length for each 2-D simulation was
chosen such that the probe envelope fluctuations induced by
filamentation did not exceed 10\% of the probe intensity, leading to
pump pulse durations of 11.4 ps, 3.8 ps and 1.1 ps for $I_0 =
10^{14}$, $10^{15}$ or $10^{16}$ W cm$^{-2}$) respectively. Results
are shown in Figure \ref{fig:1}. The top row shows the 2-D intensity
envelopes of the amplified pulses, while the bottom row shows
longitudinal and transverse intensity profiles. The 2-D plots reveal
that there is no reduction of the probe pulse diameter, allowing
amplification to high total powers, not just high intensities. The
intensity envelopes are very smooth, with hardly any fluctuations
caused by filamentation or Raman forward scattering. This is in
contrast to the results of Refs. \cite{weber13,riconda13}, which are
strongly modulated by filamentation and Raman forward scattering and
exhibit a fourfold reduction in spot diameter. Filamentation usually
occurs when either the pulse intensities are too high or the
interaction length is too long; a typical example of out-of-control
filamentation, for a pump pulse at $10^{16}$ W cm$^{-2}$ and 2 ps
duration, is shown in Figure \ref{fig:2}(a).

We define the \emph{compression ratio} as the duration of the pump
pulse divided by the duration of the amplified probe, and the
\emph{amplification ratio} as the intensity of the amplified probe
divided by the intensity of the pump. We then find compression ratios
of 40, 60 and 72, and amplification ratios of 24, 56 and 70, for pump
intensities of $10^{16}$, $10^{15}$ and $10^{14}$ W cm$^{-2}$
respectively. The increase in these ratios with decreasing pump
intensity follows from the fact that the filamentation growth rate
scales faster with pulse intensity than the strong-coupling Brillouin
scattering growth rate (see below), so using lower pulse intensities
allows one to use relatively longer interaction lengths. Of course,
using a longer interaction distance may lead to increased premature
Brillouin backscattering of the pump before it meets the probe,
potentially causing the amplified probe to have a significant
prepulse. However, we have shown elsewhere \cite{kathryn_coll} that
such premature scattering is strongly damped by collisions, and more
so for lower pump intensities that are closer to the collisional
threshold for Brillouin scattering. The pump-probe interaction itself
is well above this threshold, and therefore much less affected by
collisional damping.

We find that the absolute duration of the amplified probe increases
with decreasing pulse intensity, as follows from Eq.
(\ref{eq:ssbril}), emphasizing that Brillouin amplification works best
for longer pulses at lower intensities. The main peak of the amplified
pulse is followed by a sequence of secondary peaks, as predicted by
one-dimensional theory and simulations \cite{weber1,lehmann13a,lehmann13b}.
The amplified pulses have a ``bowed'' shape, as also seen for Raman
amplification \cite{fraiman02,trines10,trines11}. This can easily be
explained from the self-similar theory: the pump intensity is highest
on-axis and decreases for larger radius, so the probe duration is
shortest on-axis and increases for larger radius, leading to the
characteristic horseshoe shape. The energy transfer efficiency is
found to be about 30\% for each case.

Since filamentation is the most important limiting factor to Brillouin
amplification at $n_0/n_{cr} = 0.3$, it has been proposed to reduce
filamentation by lowering the plasma density to $n_0/n_{cr} = 0.05$
\cite{weber13,riconda13}. However, stimulated Raman scattering is
possible at this density, and can be expected to interfere with the
amplification process. We carried out a single 1-D static-window
simulation at $n_0/n_{cr} = 0.05$ and a plasma column length of 0.8
mm, using pulse intensities of $10^{16}$ W/cm$^2$ and a pump pulse
FWHM duration of 2.7 ps, to study the influence of Raman backward and
forward scattering on Brillouin amplification; results are displayed
in Figure \ref{fig:2}(b). Raman backscattering (RBS) was found to
generate a large prepulse to the growing probe pulse, spoiling its
contrast, while Raman forward scattering (RFS) causes the probe pulse
envelope to be strongly modulated, making RFS about as dangerous as
filamentation. A Fourier analysis of the $k$-spectrum of the pulses,
shown in Fig. \ref{fig:2}(b2) and (b3), reveals that the pump pulse
mostly suffers from Raman backward scattering, while Raman forward
scattering is dominant in the probe pulse. A close inspection of all
Raman scattering occurring during Brillouin amplification found that
the growth of the probe pulse saturates due to high levels of Raman
forward scattering, rather than Raman backscattering. If the level of
RFS in the probe pulse becomes non-linear, the coherence of the probe
pulse's carrier wave, and thus the coupling between pump and probe, is
lost, and probe amplification stops; this can be seen in Figure
\ref{fig:2}(b1). Since $\gamma_{RBS} \propto a_0 \sqrt{\omega_0
  \omega_p}$ while $\gamma_{RFS} \propto a_0 \omega_p^2 /\omega_0$, it
follows that growth of RFS and the saturation of the probe pulse are
strongly affected by the plasma density, and that lowering this
density even further, e.g. to $n_0/n_{cr} = 0.01$, will immediately
improve the pump-to-probe amplification ratio and energy transfer, see
Figure \ref{fig:2}(c). This is mainly due to a reduction in Raman
forward scattering: the ramp profile lowers the average density,
reducing the RFS growth rate and delaying the saturation of Brillouin
scattering. This effect justifies the use of a density ramp to lower
the average plasma density and obtain higher amplification gains, as
has been observed in Ref. \cite{riconda13}.  From this we conclude
that Brillouin amplification should be conducted at densities for
which RFS is either impossible ($n_0/n_{cr} > 0.25$) or unimportant
($n_0/n_{cr} \leq 0.01$). For $0.01 < n_0/n_{cr} < 0.25$, the
disadvantage of increased pump RBS and probe RFS is more serious than
the advantage of reduced probe filamentation.

As shown by Andreev \emph{et al.} \cite{weber1}, the Brillouin
amplification process is subject to the following scaling laws:
$a_{pr}(t) \propto (a_0^2 t)^{3/4}$ and $\tau_{pr}(t) \propto (a_0^2
t)^{-1/2}$, where $a_0$ is the pump amplitude and $t=\tau_{pu}/2$ the
interaction time. For high plasma densities, where Raman scattering is
not possible, the scaling laws can be extended as follows. For the
filamentation of the probe pulse, we have $\gamma_f \propto a_{pr}^2$,
so $\int \gamma_f dt \propto a_0^3 t^{5/2}$. We can keep the level of
filamentation, and thus $\int \gamma_f dt$ constant by choosing
$\tau_{pu} \propto I^{-3/5}$, where $I$ denotes the pump
intensity. This leads to $\tau_{pr}(t) \propto I^{-1/5}$ and $I_{pr}
\propto a_{pr}^2(t) \propto I^{3/5}$. Thus, the compression and
amplification ratios both scale as $\tau_{pu}/\tau_{pr} \propto
I_{pr}/I \propto I^{-2/5}$ (under the assumption that the efficiency
is mostly constant). Finally, we find that the pump pulse energy
fluence scales as $F \propto I\tau_{pu} \propto I^{2/5}$. All these
scalings are subject to the assumption that one is operating in the
strong-coupling regime for Brillouin scattering, $a_0^2 > 4(v_T/c)^3
(n_{cr}/n_0) \sqrt{1-n_0/n_{cr}} \sqrt{Zm_e/m_i}$ or $I_{pu} >
1.6\times 10^{13}$ W cm$^{-2}$ for our parameters. Already it was
found that for $I_{pu} = 10^{14}$ W cm$^{-2}$, the growing probe did
not fully conform to the above scaling laws because $I_{pu}$ is too
close to the strong-coupling threshold. Lowering the ion temperature
from 500 to 50 eV appears to lower the strong-coupling threshold
also, bringing the behaviour of the $I_{pu} = 10^{14}$ W cm$^{-2}$
case closer to pure strong-coupling Brillouin amplification and
improving its amplification and compression ratios. While ion
wave breaking has been observed in one-dimensional simulations
\cite{weber1}, with a characteristic time of $\tau_{wb} \propto
I^{-1/2}$ \cite{forslund,huller}, it did not play a major role in the
two-dimensional simulations presented above, since filamentation
always emerged earlier for pump intensities in the strong-coupling
regime. From this, it is clear that, when the pump intensity is
decreased, Brillouin amplification improves on all fronts.

Previous attempts to study Brillouin amplification in
multi-dimensional simulations \cite{weber13,riconda13} failed to
identify the correct parameter regime for optimal amplification. In
Ref. \cite{weber13}, the parameter regime for the initial probe
duration $\tau_1$ and the plasma density $n_0$ is defined as
$\tau_{sc} < \tau_1 < \tau_{wb}$ and $n_0/n_{cr} \sim 0.05$, where
$\tau_{sc} = 1/\gamma_B$ and $\tau_{wb}$ is the wave-breaking time for
the ion-acoustic wave \cite{forslund,huller}. However, a numerical
evaluation of $\tau_{sc}$ and $\tau_{wb}$ in Fig. 1 of
Ref. \cite{weber13} reveals that $\tau_{wb} < \tau_{sc}$, so the two
conditions $\tau_{sc} < \tau_1$ and $\tau_1 < \tau_{wb}$ can never be
fulfilled simultaneously and the parameter window is empty, while a plasma
density of $n_0/n_{cr} = 0.05$ is the worst possible in terms of
parasitic Raman scattering (see Fig. \ref{fig:2} above). The
ultrashort pulses presented in Refs. \cite{weber13,riconda13} do not
actually amplify: while their intensity increases by a factor 15,
their spot diameter decreases by a factor 4, so their power remains
the same. This can be explained by relativistic
self-focusing, enhanced by the presence of the pump pulse
\cite{chen-cross,shvets-cross}. Also, it is shown in Fig. 3b of
Ref. \cite{riconda13} that switching off ion motion makes no
significant difference to the intensity gain of these ultra-short
pulses, thus ruling out Brillouin amplification (which requires the
presence of an ion wave) as a contributing factor to
the pulse evolution. For the various longer-pulse cases discussed in
Refs. \cite{weber13,riconda13}, we find that $n_0$, $I_{pu}$ and
$a_1^2 \tau_{pr}^3$ are all kept constant, so these cases represent
various stages of a single configuration, rather than independent
configurations, and therefore do not constitute a true parameter scan.

In conclusion, we have studied strong-coupling Brillouin amplification
of short ($\sim 0.1$ ps) laser pulses in plasma. Amplification factors
of up to 40 have been obtained for moderate pump intensities
($10^{14}$ W cm$^{-2}$) and high plasma densities
($n_0/n_{cr}=0.3$). Notable achievements of this paper constitute: (i)
the self-similar equations (\ref{eq:ssbril}) and (\ref{eq:topt}),
which govern the evolution of the growing probe versus the pump pulse
intensity, the plasma density and the interaction length; (ii) the
identification of the plasma density and the pump pulse intensity as
the free parameters of the problem, while the initial probe pulse
duration is a dependent parameter; (iii) the use of multi-dimensional
simulations to Brillouin-amplify pulses to high power, while
preserving pulse quality and contrast, where previous work mostly
focused on high intensity and ignored pulse quality; (iv) the
identification of filamentation and probe RFS as the critical limiting
instabilities; (v) a study of parsitic Raman back- and forward
scattering in Brillouin amplification for $n_e/n_{cr} < 0.25$,
highlighting their deleterious influence on the quality and contrast
of the amplified probe pulse; (vi) identification of the following
parameter regime for efficient, high-quality Brillouin amplification:
$I_\mathrm{pump} < 10^{15}$ W cm$^{-2}$ and $n_e/n_{cr} > 0.25$, with
$n_e/n_{cr} \leq 0.01$ as an alternative; (vii) scaling laws for the
various probe pulse parameters after amplification, showing that
Brillouin amplification improves on all fronts when the pump pulse
intensity is lowered. Together, these results show that, for the right
laser-plasma configurations, Brillouin amplification is a robust and
reliable way to compress and amplify picosecond laser pulses in
plasma, and provide a comprehensive guide for the design and execution
of future Brillouin amplification experiments.

This work was supported financially by STFC and EPSRC, by the European
Research Council (ERC-2010-AdG Grant 167841) and by FCT (Portugal)
grant No. SFRH/BD/75558/2010. We would like to thank R. Kirkwood and
S. Wilks for stimulating discussions. We acknowledge PRACE for
providing access to SuperMUC based in Germany at the Leibniz research
center. Simulations were performed on the Scarf-Lexicon Cluster (STFC
RAL) and SuperMUC (Leibniz Supercomputing Centre, Garching, Germany).

\end{document}